\documentclass[aps,pre,twocolumn,showpacs,floatfix]{revtex4}
\usepackage{amssymb,amsmath,epsfig,graphicx}

\usepackage{caption}
\usepackage{dcolumn}

\newcommand{\vmedia}{\langle \dot x_{\text{cm}}\rangle}  

\newcommand{\ton}{t_{\text{on}}}
\newcommand{\toff}{t_{\text{off}}}

\newcommand{\Con}{C_{\text{on}}}
\newcommand{\Coff}{C_{\text{off}}}
\newcommand{\lambdaon}{\lambda_{\text{on}}}
\newcommand{\lambdaoff}{\lambda_{\text{off}}}

\newcommand{\taumin}{\tau_{\text{min}}}

\newcommand{\be}{\begin{equation}}
\newcommand{\ee}{\end{equation}}

\begin{document}

\title{\bf Flux enhancement and multistability induced by time delays in
a feedback controlled flashing ratchet}

\author{F. J. Cao}
\email{francao@fis.ucm.es}
\affiliation{Departamento de F\'{\i}sica At\'omica, Molecular y
Nuclear,
Universidad Complutense de Madrid, \\
Avenida Complutense s/n, 28040 Madrid, Spain.}
\affiliation{LERMA, Observatoire de Paris, Laboratoire Associ\'e au CNRS UMR 8112, \\
61, Avenue de l'Observatoire, 75014 Paris, France.}

\author{M. Feito}
\email{feito@fis.ucm.es}
\affiliation{Departamento de F\'{\i}sica At\'omica, Molecular y Nuclear,
Universidad Complutense de Madrid, \\
Avenida Complutense s/n, 28040 Madrid, Spain}


\begin{abstract}
Feedback controlled ratchets are thermal rectifiers that use information
on the state of the system to operate. We study the effects of \emph{time
  delays} in the feedback for a protocol that performs an instantaneous 
maximization of the
center-of-mass velocity in the \emph{many particle} case. For
small delays the center-of-mass velocity decreases for increasing
delays (although not as fast as in the few particle case).
However, for large delays we find the surprising result that the
presence of a delay can improve the flux performance of the
ratchet. In fact, the maximum flux obtained with the optimal periodic
protocol is attained. This implies that the delayed feedback protocol
considered can perform better than its non-delayed counterpart.
The improvement of the flux observed in the presence of large delays is
the result of the emergence of a new dynamical regime where the
presence of the delayed feedback stabilizes quasiperiodic
solutions that resemble the solutions obtained in a certain
closed-loop protocol with thresholds. In addition, in this new regime the
system presents multistability, i.e. several quasiperiodic solutions can be
stable for a fixed time delay. 

\end{abstract}


\maketitle

\section{Introduction}

Brownian motors or ratchets are mechanisms that induce transport
rectifying the motion of Brownian particles through the
introduction of a time-dependent perturbation that drives the
system out of equilibrium \cite{rei02}. This type of systems allow to get
insight in non-equilibrium processes \cite{rei02}. In addition,
they are also important due to their applications to many fields as
nanotechnology and biology~\cite{rei02,lin02}.

The two main types of ratchets are rocking ratchets~\cite{mag93,ast94} and
flashing ratchets~\cite{ajd93,ast94}. In rocking ratchets (also called tilting
ratchets) the perturbation acts as a time-dependent additive
driving force, which is unbiased on the average, while in flashing
ratchets (also called pulsating ratchets) the time-dependent
perturbation changes the potential shape without affecting its
spatial periodicity. An example of a flashing ratchets is a ratchet that
operates switching on and off a spatially periodic asymmetric
potential. In this particular case it can be seen that a simple
periodic or random switching can rectify thermal fluctuations and
produce a net current of particles.

A new class of ratchets that use information on the state of the
system to operate have been introduced in Ref.~\cite{cao04}. These
\emph{feedback ratchets} (or closed-loop ratchets) are able to increase
the net current and the power output of collective Brownian
ratchets~\cite{cao04,din05,fei06,fei07}. Feedback can be
implemented monitoring the positions of the particles (see for
example Refs.~\cite{rou94,mar02}) and subsequently using the
information gathered to decide whether to switch on or off the ratchet
potential according to a given protocol. In addition, feedback ratchets have
been recently suggested as a mechanism to explain the stepping motion of the
two-headed kinesin~\cite{bie07}. 

The first feedback protocol proposed was the so-called instantaneous
maximization of the center-of-mass velocity~\cite{cao04}, which switches on
the potential only if switching on would imply a positive
displacement for the center-of-mass position (i.e., if the net force with the
potential on would be positive). The instantaneous maximization protocol gives
the maximum current in the case of one 
particle and performs better than any open-loop protocol for few
particles. However, it has a very low performance in the many
particle case given an average center-of-mass velocity smaller
than that obtained with an optimal periodic protocol. (We call
many particle case the case when the fluctuations of the net force
are smaller than its maximum absolute value.) An improvement of
the instantaneous maximization protocol is the threshold
protocol~\cite{din05}, which consist on introducing two
threshold values in order to switch the potential before the net
force reaches a zero value. In this way, there is an increase of the
performance for many particles up to velocity values equaling the ones of
the optimal open-loop periodic protocol. 

In order to check if it is experimentally feasible to obtain the
increase of performance theoretically predicted for the few
particle case one important question is to check the effects of
time delays in the feedback that would be present in any experimental 
implementation. These time delays in the feedback come from the fact
that the measure, transmission, processing, and action steps take
a finite time interval \cite{ste94,bec05}. Time delays in the feedback also 
appear 
naturally in complex systems with self regulating mechanisms
\cite{boc00,fra05b}. Recently, we have investigated the effects that
the delay has in the operation of feedback controlled ratchets
in the \emph{few particle} case~\cite{fei07b}. We have found that
even in the presence of time delays feedback controlled ratchets
can give better performance than the corresponding optimal
open-loop ratchet, although time delays decrease the performance.

In this paper we investigate the effects of time delays in the
instant maximization protocol for the
\emph{many particle} case. We find that for small delays the
asymptotic average center-of-mass velocity decreases for
increasing delays (although not as fast as in the few particle case).
However, if we continue increasing the time delay the average velocity
starts to increase up to the value obtained for an optimal open-loop
protocol. This surprising result makes that for many particles the instant 
maximization protocol gives greater average velocities in the
presence of delay than in its absence. In Sec.~\ref{sec:model}
we present the evolution equations of the system. In the next
section, Sec.~\ref{sec:results}, we briefly review the results for 
zero delays that will be
useful, and thereafter we expose the results in the two dynamical
regimes: small delays and large delays. Finally, in
Sec.~\ref{sec:con} we summarize and discuss the results.

\section{The model} \label{sec:model}

The feedback ratchet we consider consists of $ N $ Brownian
particles at temperature $T$ in a periodic potential $ V(x) $. The
force acting on the particles is $ F(x) = - V'(x) $, where the
prime denotes spatial derivative. The state of this system is
described by the positions $x_i(t)$ of the particles satisfying
the overdamped Langevin equations
\begin{equation}\label{langevin}
\gamma \dot x_i(t)=\alpha(t)F(x_i(t))+\xi_i(t);\quad i=1,\dots,N,
\end{equation}
where $\gamma$ is the friction coefficient (related to the
diffusion coefficient $D$ through Einstein's relation
$D=k_BT/\gamma$) and $\xi_i(t)$ are Gaussian white noises of zero
mean and variance $\langle \xi_i(t)\xi_j(t^\prime)\rangle =2\gamma
k_B T\delta_{ij}\delta(t-t^\prime)$. The control policy uses the
sign of the net force per particle,
\begin{equation}
f(t)=\frac{1}{N}\sum_{i=1}^N F(x_i(t)),
\end{equation}
as follows: The controller measures the sign of the net force and,
after a time $ \tau $, switches the potential on ($\alpha=1$) if
the net force was positive or switches the potential off
($\alpha=0$) if the net force was negative. Therefore, the delayed
control protocol considered is
\begin{equation}\label{alfa-delay}
  \alpha(t)=\Theta(f(t-\tau)),
\end{equation}
with $\Theta$ the Heaviside function [$\Theta (x)=1$ if $x>0$,
else $\Theta (x)=0$].

\begin{figure}
  \begin{center}
    \includegraphics [scale=0.5] {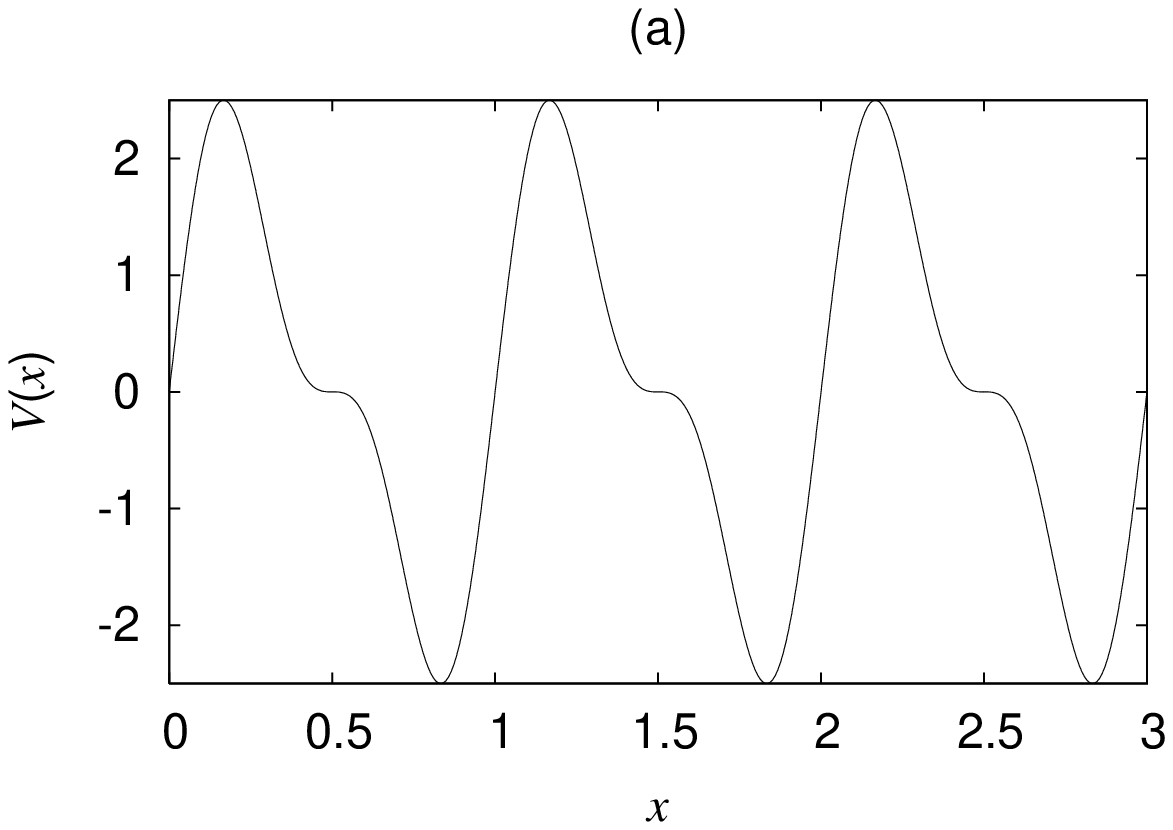}
    \includegraphics [scale=0.5] {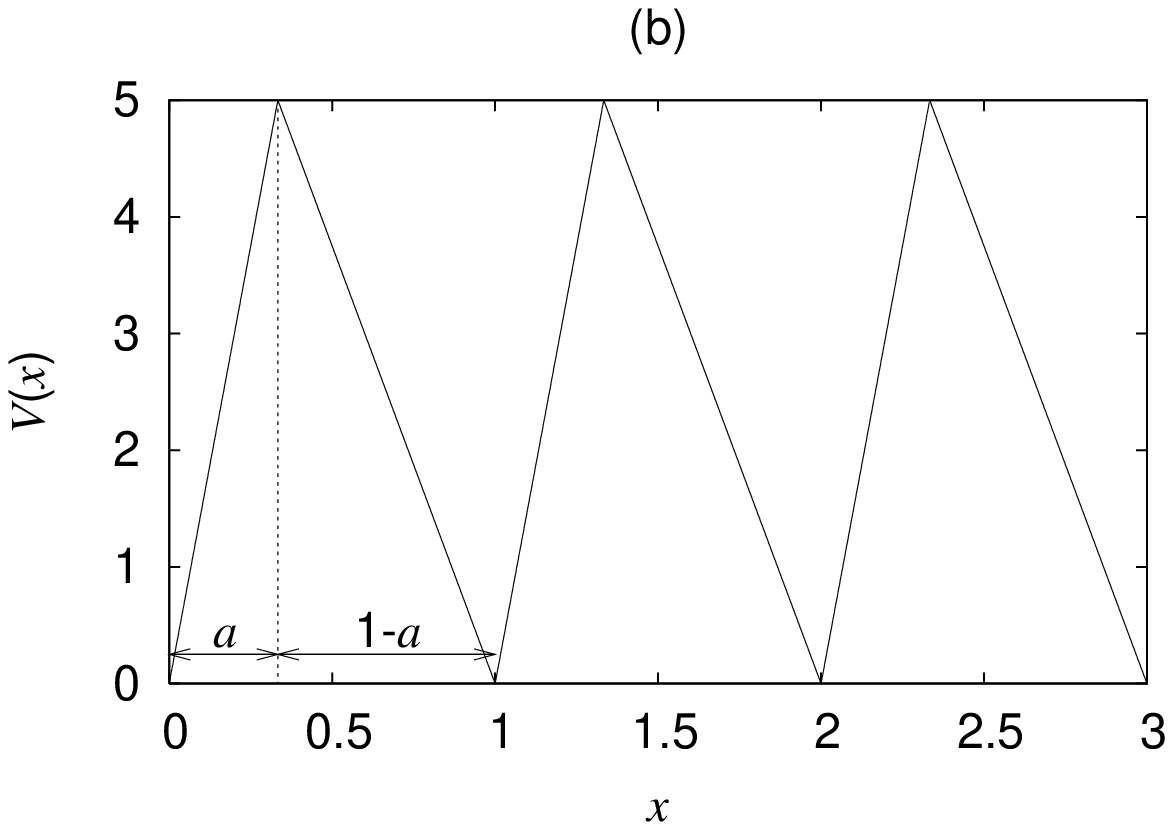}
  \end{center}
  \caption{Panel (a): `Smooth' potential \eqref{smoothpot} for $V_0=5k_BT$.
    Panel (b): `Saw-tooth' potential \eqref{sawtoothpot} for $V_0=5k_BT$ and $
  a = 1/3$. Units: $L=1$, $k_BT=1$.
  }
  \label{fig:pot}
\end{figure}

As ratchet potential we consider the `smooth' potential of period
$L$ and height $V_0$ (Fig.~\ref{fig:pot}~a),
\be \label{smoothpot}
V(x) = \frac{2V_0}{3\sqrt{3}} \left[ \sin \left(\frac{2\pi x}{L}
\right) + \frac{1}{2} \sin \left( \frac{4\pi x}{L} \right)
\right].
\ee
We have also verified that analogous results are obtained for the
`saw-tooth' potential of period $L$, i.e. $V(x)=V(x+L)$, height
$V_0$, and asymmetry parameter $a<1/2$
(Fig.~\ref{fig:pot}~b),
\begin{equation} \label{sawtoothpot}
  V(x)=
  \begin{cases}
    \frac{V_0}{a}\frac{x}{L} &\text{if } 0\leq \frac{x}{L}\leq a ,\\
    V_0-\frac{V_0}{1-a}\left(\frac{x}{L}-a\right) &\text{if } a<
    \frac{x}{L}\leq 1.
  \end{cases}
\end{equation}
The height $V_0$ of the potential is the potential difference
between the potential at the minimum and at the maximum, while $aL$ is
the distance between the minimum and the maximum positions. In
view of this definition, the `smooth' potential \eqref{smoothpot}
has asymmetry parameter $ a= 1/3 $.

Throughout the rest of this paper, we will use units where $L=1$,
$k_BT=1$, and $D=1$.

We consider in this paper the \emph{many particles} case that is
characterized by the fact that the typical fluctuations of the net
force are smaller than the maximum values of its absolute value.

\section{Delayed many particle feedback ratchet}\label{sec:results}

We study the effects of time delays in the previous feedback
controlled Brownian ratchets in the many particles case,
considering both the `smooth' potential and the `saw-tooth'
potential for various potential heights and different initial
conditions.

\begin{figure}
  \begin{center}
    \includegraphics [scale=0.6] {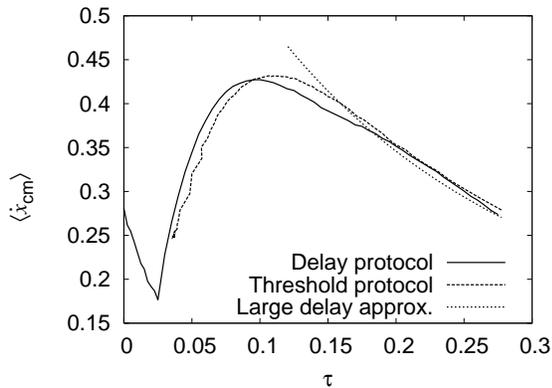}
  \end{center}
  \caption{Center-of-mass velocity as a function of the delay (for large
  delays only the first branch is represented here), and comparison with the
  results obtained with the threshold protocol and with the large delay
  approximation Eq.~\eqref{vlargetau}. For the `smooth' potential
  \eqref{smoothpot} with $V_0=5k_BT$ and $ N = 10^5 $ particles.
  Units: $L=1$, $D=1$, $k_BT=1$.
  }
  \label{fig:taumin}
\end{figure}

We find that the system presents two regimes separated by a delay
$\taumin$ for which the center-of-mass velocity has a minimum; see
Fig.~\ref{fig:taumin}. In the small delay regime ($ \tau < \taumin
$) the flux decreases with increasing delays as one could expect.
On the contrary, in the large delay regime ($ \tau > \taumin $) we
have observed and explained a surprising effect, namely, the
center-of-mass velocity increases for increasing delays and the
system presents several stable solutions. We have found that this
critical time delay $ \taumin $ is inversely proportional to the
potential height $ \taumin \propto 1/V_0 $ with a proportionality
constant that mildly depends on the number of particles.

\subsection{Zero delay}\label{sec:zero}

The many particle ratchet in absence of delay (i.e., $\tau=0$ in
the model of Sec.~\ref{sec:model}) have been studied in
Ref.~\cite{cao04}. It has been shown that the net force per
particle exhibits a quasideterministic behavior that alternates
large periods of time $t_{\text{on}}$ with $ f(t)>0 $ (on
dynamics) and large periods of time $t_{\text{off}}$ with $ f(t)<0
$ (off dynamics). The center-of-mass velocity can be computed as
\be \label{vmediacero}
\vmedia = \frac{\Delta x(\ton)}{\ton+\toff},
\ee
with
\be \label{deltaxon}
\Delta x(\ton) = \Delta x_{\text {on}} [
1-e^{-\ton/(2\Delta t_{\text {on}})} ],
\ee
where $\Delta x_{\text{on}}$ and $ \Delta t_{\text{on}} $ are
obtained fitting the displacement during the `on' evolution for an
infinite number of particles (see Ref.~\cite{fei06} for details).

On the other hand, for many particles the fluctuations of the net
force are smaller than the maximum value of the net force.
This allows the decomposition of the dynamics as the dynamics for
an infinite number of particles plus the effects of the
fluctuations due to the finite value of $N$. The late time
behavior of the net force $f(t)$ for an infinite number of
particles is given for the on and off dynamics by \cite{cao04},
\be
f_\nu^\infty(t) = C_\nu e^{-\lambda_\nu(t-\tau_\nu)} \mbox{ with }
\nu = \mbox{on, off}.
\ee
The coefficients $ C_\nu $, $\lambda_\nu$, and $\tau_\nu$ can be
obtained fitting this expression with the results obtained
integrating a mean field Fokker-Planck equation obtained in the
limit $ N \to \infty $ and without delay; see
Refs.~\cite{cao04,fei06} for details. For a finite number of
particles the fluctuations in the force induce switches of the
potential and the times on and off are computed equating $
f^\infty_\nu $ to the amplitude of the force fluctuations, resulting
\cite{cao04}
\be \label{tonoffzero}
\ton + \toff = b + d \ln N,
\ee
with $ b = \Con + \Coff $ and $ d = (\lambdaon + \lambdaoff) /
(2\lambdaon\lambdaoff) $.

\subsection{Small delays}\label{sec:small}

\begin{figure}
  \begin{center}
    \includegraphics [scale=0.6] {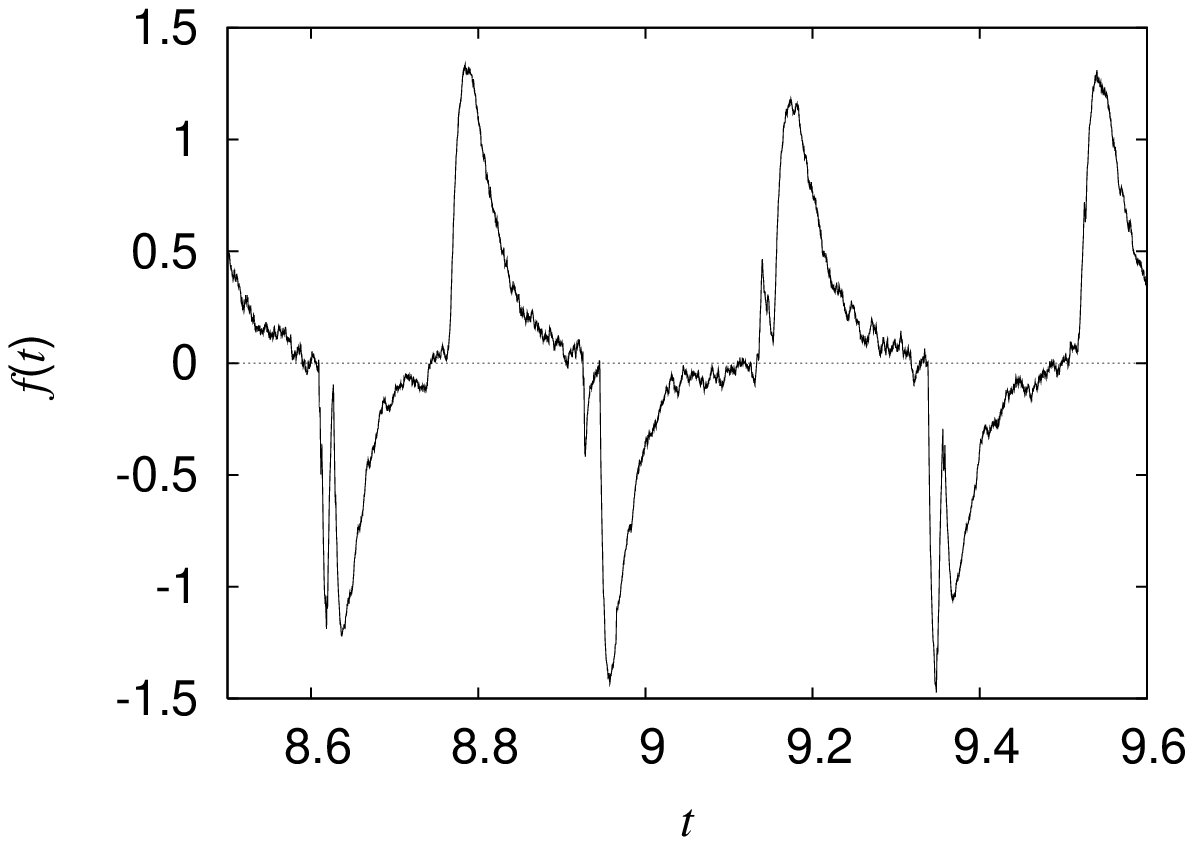}
  \end{center}
  \caption{Evolution of the net force with a small delay ($\tau=0.02$)
  for the `smooth' potential
  Eq.~\eqref{smoothpot} with  $V_0=5k_BT$ and
  $N=10^5$ particles. Units: $L=1$, $D=1$, $k_BT=1$.
  }
  \label{fig:evol}
\end{figure}

For small delays, $ \tau < \taumin $, we observe that the flux
decreases with the delay. See Fig.~\ref{fig:taumin}. We have seen
that this decrease is slower than that found for the few particle
case~\cite{fei07b}, and that the expressions derived to describe this decrease
in the few particles case does not hold here.
However, the decrease observed here can be understood by the fact
that the delay implies an increase of the time interval between
switches, which makes the tails of $ f(t) $ longer than for no
delay and the form of $ f(t) $ less smooth than for no delay. See
Fig.~\ref{fig:evol}. The main effect of the delay is to stretch
the `on' and `off' times of the dynamics; then, using the many
particle approximation \cite{cao04} we can write
\begin{equation}\label{small-del}
    \vmedia=\frac{\Delta x_{\text{on}}}{\ton+\toff+\Delta\tau}=
    \frac{\Delta x_{\text{on}}}{b+d \ln N +\Delta\tau},
\end{equation}
where we have found that the increase of the length of the on-off
cycle $ \Delta\tau $ is proportional to the delay $ \Delta\tau
\propto \tau$.

\subsection{Large delays}\label{sec:large}

\begin{figure}
  \begin{center}
    \includegraphics [scale=0.6] {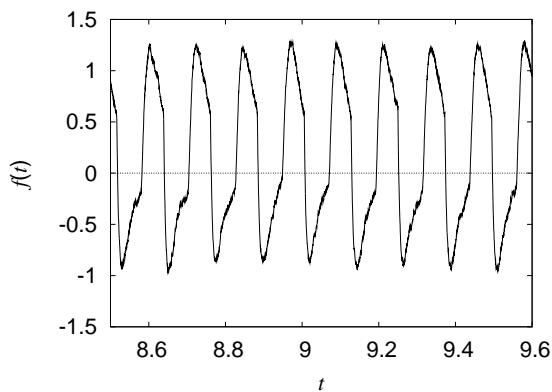}
  \end{center}
  \caption{Evolution of the net force with a large delay ($\tau=0.12$)
  for the `smooth' potential Eq.~\eqref{smoothpot} with  $V_0=5k_BT$ and
  $N=10^5$ particles. Units: $L=1$, $D=1$, $k_BT=1$.
  }
  \label{fig:evollarge}
\end{figure}

After the minimum flux is reached for $ \tau = \taumin $, the flux
begins to increase with the time delay (see
Fig.~\ref{fig:taumin}). This increase is due to a change in the
dynamical regime: for $ \tau > \taumin $ the present net force
starts to be nearly synchronized with the net force a time $ \tau
$ ago. This
\emph{selfsynchronization} gives rise to a quasiperiodic solution of
period $ T = \tau $. Note that there is not a strict periodicity due to
stochastic fluctuations in the `on' and `off' times. Looking at
the $f(t)$ dependence, Fig.~\ref{fig:evollarge}, we see that the
solutions stabilized by the selfsynchronization are similar to
those obtained with the threshold protocol \cite{din05,fei06}. In
Fig.~\ref{fig:taumin} we show that the threshold protocol that has
the same period gives similar center-of-mass velocity values,
confirming the picture. (Differences are due to the fact that we
have considered for the threshold protocol simulations with on and
off thresholds of the same magnitude, while Fig.~\ref{fig:evollarge} 
shows that the effective thresholds are different.)

This picture allows to understand the increase of velocity for
increasing delay, and the presence of a maximum. This maximum is
related with the optimal values of the thresholds that have been
shown in \cite{fei06} to give a quasiperiodic solution of period $
{\cal T}_{\text{on}} + {\cal T}_{\text{off}} $, with $ {\cal
T}_{\text{on}} $ and $ {\cal T}_{\text{off}} $ the optimal `on'
and `off' times of the periodic protocol. Therefore, if we know
the values of $ {\cal T}_{\text{on}} $ and $ {\cal T}_{\text{off}}
$ for the optimal periodic protocol [${\cal T}_{\text{on}} \sim (1-a)^2/V_0 $ 
and $ {\cal T}_{\text{off}} \sim a^2/2 $] we can predict that the
maximum of the center-of-mass velocity is reached for a delay
\be
\tau_{\text{max}} = {\cal T}_{\text{on}} + {\cal T}_{\text{off}},
\ee
and has a value
\be
\vmedia_{\text{closed}} (\tau_{\text{max}}) =
\vmedia_{\text{open}}^{\text{max}},
\ee
with $ \vmedia_{\text{open}}^{\text{max}} $ the center-of-mass
velocity for the optimal open-loop protocol. Thus, this
expression gives the position and height of the maximum of the
delayed feedback control protocol in terms of the characteristic
values of the optimal open-loop control. In particular, it implies
that the position and height of the maximum for the flux is
independent of the number of particles. 

As an example we can apply these expressions to the `smooth'
potential with $V_0 = 5$ that for the optimal periodic protocol
gives $\vmedia = 0.44$ for ${\cal T}_{\text{on}}=0.06$ and ${\cal
T}_{\text{off}}=0.05$, so we obtain  
$\tau_{\text{max}}=0.06+0.05=0.11$ in agreement with Fig.~\ref{fig:taumin}.  

\bigskip

\begin{figure}
  \begin{center}
    \includegraphics [scale=0.6] {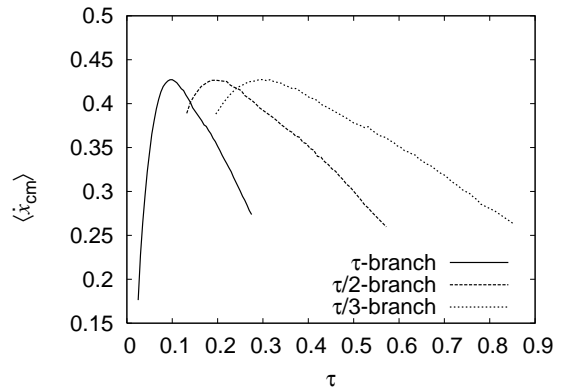}
  \end{center}
  \caption{First three branches of stable solutions
  for the `smooth' potential \eqref{smoothpot}
  with $V_0=5k_BT$ and $ N = 10^5 $ particles. Units: $L=1$, $D=1$, $k_BT=1$.
  }
  \label{fig:vbranches}
\end{figure}

For values of the delay of the order of or larger than $
\tau_{\text{max}} $ quasiperiodic solutions of other periods start
to be stable; see Fig.~\ref{fig:vbranches}. The periods for the
net force $f(t)$ that are found are those that fit an integer
number of periods inside a time interval $\tau$, verifying that
the present net force is synchronized with the net force a time
$\tau$ ago, that is, the quasiperiodic solutions have periods $ T =
\tau/2 $, $ T = \tau/3 $, $\ldots$ In addition, it can be seen
that the center-of-mass velocity of the $n$ branch
$\vmedia_{\tau/n}$ whose $f(t)$ has period $ T = \tau/n $ is
related with that of the $ T = \tau $ branch through
\be \label{vmedian}
\vmedia_{\tau/n}(\tau) = \vmedia_\tau (\tau/n).
\ee
We highlight that several branches can be stable for the same time
delay $\tau$. Whether the system finally goes to one or another
stable solution depends on the initial conditions and on the
particular realization of the noise. See Fig.~\ref{fig:vbranches}.
For these branches we have found initial conditions that goes to
these solutions and that remain in them during several thousands of periods, 
indicating that they are stable solutions or at least
metastable solutions with a large lifetime.

The analogy with the threshold protocol allows to use the analytic
results of \cite{fei06} to get further insight in the numerical
results. The behavior for large delays for the $ T=\tau $ branch
can be obtained using the relation 
\be \label{vlargetau}
\vmedia = \frac{\Delta x(\tau)}{\tau},
\ee
with $ \Delta x(\tau) $ given by Eq.~\eqref{deltaxon}. This
equation gives a good prediction for the largest delays of the
first branch (see Fig.~\ref{fig:taumin}).

On the other hand, for very large values of the delays of the first
branch the solutions in a given branch start to become unstable,
what can be understood noting that this happens when the
fluctuations of the net force become of the order of the absolute
value of the net force. Thus, the maximum delay that gives a
stable solution in the first branch is
\be \label{tauinst}
\tau_{\text{inst}} = \ton+\toff = b+d \ln N,
\ee
where $b$ and $d$ are determined as in Eq.~\eqref{tonoffzero}. For
example, for the `smooth' potential with $ V_0 = 5 $, which has $ b =
-0.070 $ and $d=0.031$, we obtain for $N = 10^5$
particles the value $\tau_{\text{inst}}=0.29$ in accordance with the
numerical results shown in Figs.~\ref{fig:taumin} and
\ref{fig:vbranches}.

The previous results for the first branch, Eqs.~\eqref{vlargetau}
and \eqref{tauinst}, can be extended to other branches by direct
application of the relation \eqref{vmedian}.

\section{Conclusions}\label{sec:con}

We have studied the effects of time delays in the
many particle case, where surprising and interesting results
arise. Although in the many particle case without delay the
instantaneous maximization protocol performs worst than the
optimal open-loop protocol, the introduction of a delay can
increase the center-of-mass velocity up to the values given by the
optimal open-loop control protocol.

For small delays the asymptotic average velocity decreases for
increasing delays, until it reaches a minimum. 
After this minimum, a change of regime happens and the
system enters a selfsynchronized dynamics with the net force at
present highly correlated with the delayed value of the net force
used by the controller. This selfsynchronization stabilizes
several of the quasiperiodic solutions that can fit an integer
number of periods in a time interval of the length of the time
delay. The stable quasiperiodic solutions have an structure
similar to those solutions appearing in the threshold protocol.
This analogy has allowed us to make numerical and analytical
predictions using the previous results for the threshold protocol
\cite{fei06}. In particular, we have established the location and value of
the maximum, and also the value of the time delay beyond
which a quasiperiodic solution becomes unstable. The results
obtained shown that for most time delays several solutions are
stable and therefore the systems presents multistability; which stable
solution is reached depends on the past history of the 
system.

The possibility to choose the quasiperiod of the solution we want
to stabilize just tuning the time delay can have potential
applications to easily control the particle flux. Note that we can
even leave some branch just going to time delays where the branch
is already unstable, and force the system to change to another
branch of solutions.

\acknowledgments

We acknowledge financial support from the MEC (Spain) through
Research Projects FIS2005-24376-E and FIS2006-05895, and from the
ESF Programme STOCHDYN. In addition, MF thanks the Universidad
Complutense de Madrid (Spain) for support through grant ``Beca
Complutense''.

\end{document}